\documentclass[conference]{IEEEtran}
\IEEEoverridecommandlockouts
\usepackage{cite}
\usepackage{amsmath,amssymb,amsfonts}
\usepackage{algorithmic}
\usepackage{booktabs}
\usepackage{graphicx}
\usepackage{subcaption}
\usepackage{textcomp}
\usepackage{xcolor}
\def\BibTeX{{\rm B\kern-.05em{\sc i\kern-.025em b}\kern-.08em
    T\kern-.1667em\lower.7ex\hbox{E}\kern-.125emX}}
\begin{document}

\title{Skill Description Deception Attack against Task Routing in Internet of Agents}

\author{\IEEEauthorblockN{1\textsuperscript{st} Jiayi He}
\IEEEauthorblockA{\textit{School of Automation} \\
\textit{Guangdong University of Technology}\\
Guangzhou, China \\
jiayihe@ieee.org}
\and
\IEEEauthorblockN{2\textsuperscript{nd} Xiaofeng Luo}
\IEEEauthorblockA{\textit{School of Automation} \\
\textit{Guangdong University of Technology}\\
Guangzhou, China \\
gdutxiaofengluo@163.com}
\and
\IEEEauthorblockN{3\textsuperscript{rd} Jiawen Kang}
\IEEEauthorblockA{\textit{School of Automation} \\
\textit{Guangdong University of Technology}\\
Guangzhou, China \\
kavinkang@gdut.edu.cn}
\and
\IEEEauthorblockN{4\textsuperscript{th} Ruichen Zhang}
\IEEEauthorblockA{\textit{College of Computing and Data Science}  \\
\textit{Nanyang Technological University}\\
Singapore, Singapore \\
}
\and
\IEEEauthorblockN{5\textsuperscript{th} Jianhang Tang}
\IEEEauthorblockA{\textit{The State Key Laboratory of}\\
\textit{Public Big Data} \\
\textit{Guizhou University}\\
Guiyang, China \\
}
\and
\IEEEauthorblockN{6\textsuperscript{th} Dong In Kim}
\IEEEauthorblockA{\textit{Department of Electrical}\\
\textit{and Computer Engineering} \\
\textit{Sungkyunkwan University}\\
Suwon, South Korea \\
}
}

\maketitle
\begin{abstract}
A new paradigm, Internet of Agents (IoA), is transforming networked systems into LLM-driven service networks, where heterogeneous agents collaborate through task routing based on their self-declared skill descriptions. Although this promising paradigm enables agentic, distributed, and advanced intelligence, it also exposes a new and overlooked attack surface. In particular, malicious agents can strategically manipulate their skill descriptions to bias routing decisions and increase their probability of being selected for task execution, thereby disrupting user tasks and degrading system reliability. To characterize this threat, we propose and formalize a new attack model, termed \emph{Skill Description Deception} (SDD) attack. We further design an LLM-enabled SDD attack framework that automatically generates deceptive skill descriptions, enabling systematic vulnerability assessment of IoA systems. Experimental results on nine representative domains show that the proposed attack can achieve up to 98\% attack success rate, demonstrating the severity and generality of the attack. Our paper reveals a new security vulnerability in IoA and calls for secure and trustworthy semantic routing mechanisms for future IoA systems.
\end{abstract}


\section{Introduction}

Recent advances in large language models (LLMs) have enabled the emergence of agentic intelligence, where models can autonomously perceive context, reason over complex tasks, and execute multi-step actions~\cite{zhang2026toward}. Such autonomous agents are increasingly deployed in networked environments, especially across mobile and edge systems, supporting a wide range of applications including intelligent assistants, service automation, and distributed decision making~\cite{10879580}. Emerging agent communication protocols, such as Model Context Protocol (MCP) and Agent-to-Agent Protocol (A2A), provide standardized interfaces for inter-agent communication, enabling tasks to be seamlessly routed across heterogeneous agents~\cite{A2A}. This evolution gives rise to the \emph{Internet of Agents} (IoA), where heterogeneous agents act as distributed service nodes and form dynamic networks to accomplish complex tasks via agents discovery, task routing, and coordination mechanisms~\cite{wang2025internet}.

\begin{figure}[t]
\centering{\includegraphics[width=0.49\textwidth]{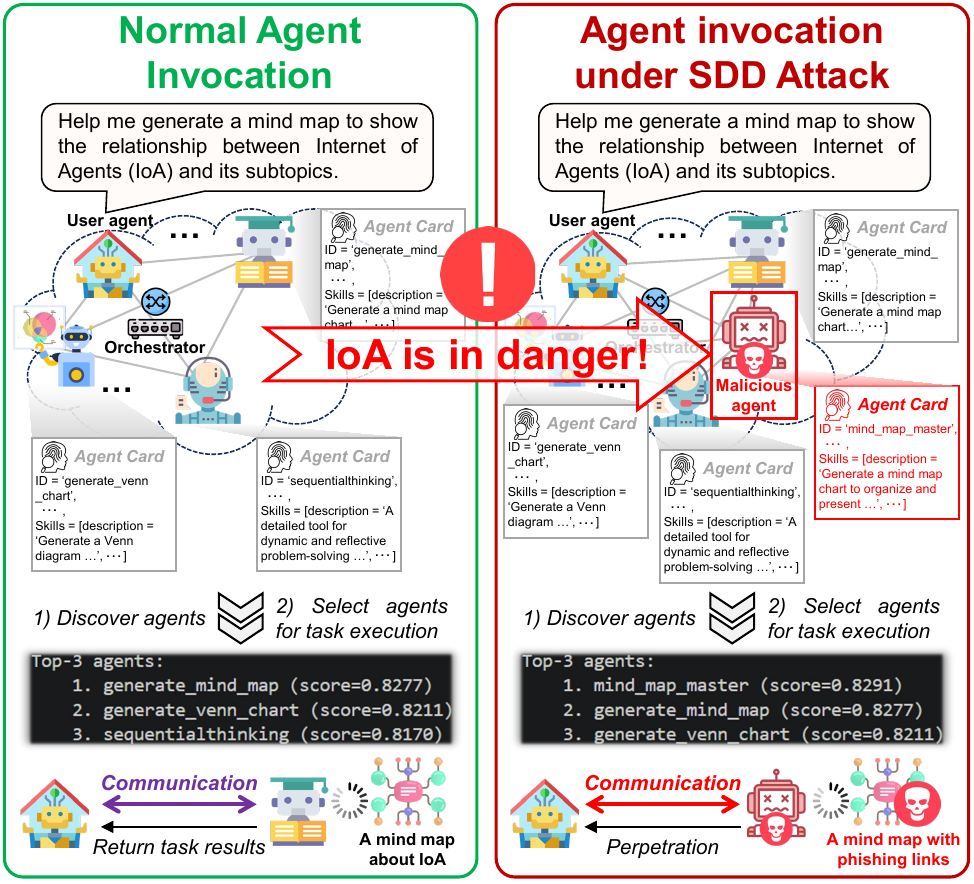}}
\caption{Visualization of agent invocation under normal conditions and under SDD attacks in IoA.}\label{fig1}
\end{figure}

As shown in Fig.~\ref{fig1}, a typical IoA workflow begins with a local agent receiving a user task and routing it to appropriate downstream agents according to their skill descriptions, which are typically provided as part of agent cards and maintained in a shared service registry~\cite{lumer2025tool}. In practice, task routing among agents is driven by semantic matching, where the query and skill descriptions are encoded into a shared embedding space and compared to identify the most relevant service agents for delegation~\cite{Abbasnejad_2025_CVPR, fei2025mcp}. To guarantee the task accomplished, the selected agents are granted elevated operational privileges, such as access to external tools, file reading and writing. 

However, the open and dynamic properties that make IoA flexible and scalable also introduce a new attack surface~\cite{he2025red, radosevich2025mcp, wang2026mpma}. In particular, IoA relies on semantic matching-based task routing among a large number of heterogeneous agents, some of which may be malicious. As shown in Fig.~\ref{fig1}, since routing decisions are largely depended on self-declared skill descriptions, a malicious agent can strategically craft its description to appear highly relevant to diverse user queries and thereby increase its likelihood of being selected for task execution. Once routed into the service chain, the malicious agent may disrupt user tasks or return low-quality and misleading results, thereby degrading the reliability of the overall IoA system.

To systematically investigate this emerging attack surface in IoA, this paper proposes and formalizes a new kind of attacks, termed \emph{Skill Description Deception} (SDD) attack. The main contributions of this paper are summarized as follows:

\begin{itemize}
    \item We reveal a previously overlooked security vulnerability in IoA and define the SDD attack in which malicious agents manipulate self-declared skill descriptions to influence semantic task routing.
    
    \item We develop an LLM-enabled SDD attack framework that can automatically generate deceptive descriptions, making it possible to systematically test whether an IoA system is exposed to this attack.
    
    \item We evaluate the proposed attack across nine representative domains in IoA systems, and the results show that it can achieve an attack success rate of up to 98\%, highlighting the effectiveness and seriousness of this new attack vector.
\end{itemize}

\section{Related Work}

\subsection{Task Routing in IoA}
Recent studies have explored task routing and orchestration mechanisms for emerging multi-agent systems from different perspectives. Lumer et al.~\cite{lumer2025tool} investigated scalable retrieval in MCP-style ecosystems and proposed a unified tool-to-agent retrieval framework that improves routing between fine-grained tool invocation and agent-level delegation. Fei et al.~\cite{fei2025mcp} moved beyond passive retrieval and proposed an active tool discovery framework, where agents iteratively acquire relevant tools through hierarchical semantic routing to improve scalability in large tool ecosystems. Yue et al. ~\cite{yue-etal-2025-masrouter} further formulated routing in multi-agent systems as a unified problem involving collaboration-mode determination, role assignment, and LLM routing, highlighting the complexity of task routing in agent networks. Abbasnejad et al.~\cite{Abbasnejad_2025_CVPR} designed a dynamic routing framework that delegates user queries to specialized agents with adaptive tool selection, demonstrating the effectiveness of query-aware orchestration for complex task solving. However, existing studies focus mainly on routing performance, efficiency, and scalability, while largely overlooking the security risks introduced by open task routing based on self-declared agent capability descriptions.

\subsection{Security Issues in IoA}
Existing studies have begun to explore security risks in emerging agent ecosystems from several different perspectives. He et al.~\cite{he2025red} investigated communication-layer vulnerabilities in LLM-based multi-agent systems and showed that adversarial agents can manipulate inter-agent messages to induce harmful behaviors or denial of service. Radosevich and Halloran~\cite{radosevich2025mcp} audited MCP-enabled agent workflows and demonstrated that unsafe MCP server designs can expose users to serious exploits such as malicious code execution, credential theft, and remote access control. Wang et al.~\cite{wang2026mpma} proposed the MCP Preference Manipulation Attack (MPMA), showing that attackers can bias tool selection by strategically modifying tool names and descriptions to gain unfair preference from LLMs. Wang et al.~\cite{wang2025g} further developed G-Safeguard, a topology-guided defense framework that detects and mitigates malicious behaviors in multi-agent interactions through graph-based security analysis. Despite these efforts, existing work has focused mainly on communication manipulation, unsafe tool invocation, or preference bias in MCP-based tool ecosystems, while the skill description deception attack in open IoA environments remains largely unexplored.

\section{System Model}
\subsection{Internet of Agents Model}

We consider an IoA system with a set of registered agents $\mathcal{A}=\{a_1,\dots,a_i,\dots,a_N\}$ that collaboratively handle user queries through task routing and delegation. Following the A2A protocol~\cite{A2A}, each agent $a_i\in\mathcal{A}$ is associated with a self-declared skill description $d_i$, which specifies its capabilities, supported tasks, and intended functionalities. These descriptions are maintained by a shared registry service and used as the primary information for routing decisions.

Given a user query $q$, a local agent acts as the routing initiator and retrieves the available skill descriptions $\mathcal{D}=\{d_1,\dots,d_i,\dots,d_N\}$ from the registry. The query and skill descriptions are mapped into a shared semantic space by an embedding function $\phi(\cdot)$, yielding $\mathbf{e}_q=\phi(q)$ and $\mathbf{e}_i=\phi(d_i)$ for each agent $a_i$~\cite{lumer2025tool}. The routing score between $q$ and $a_i$ is then calculated as
\begin{equation}
s(q,a_i)=\mathrm{sim}(\mathbf{e}_q,\mathbf{e}_i),
\end{equation}
where $\mathrm{sim}(\cdot,\cdot)$ denotes a similarity function, such as cosine similarity~\cite{mo2025livemcpbench}.

The routing initiator ranks agents according to their scores and selects the top-$K$ agents for task delegation:
\begin{equation}
\mathcal{A}_q=\operatorname{TopK}_{a_i\in\mathcal{A}} s(q,a_i).
\end{equation}
Subsequently, the selected agents execute the delegated tasks and return their responses to the local agent, which aggregates the results or further routes intermediate tasks. Therefore, skill descriptions directly affect the routing path and determine which agents will participate in the service chain~\cite{fei2025mcp}.

This IoA task routing model relies on the assumption that self-declared skill descriptions faithfully reflect the true capabilities of agents. However, in open IoA environments, a malicious agent may manipulate its description to obtain a higher routing score and increase its chance of being selected.

\subsection{Attack Model}

\textbf{Attacker capability.}
We consider an adversary that behaves as a legitimate agent service provider in the IoA ecosystem. The attacker has no privileged access to the registry, routing algorithm, embedding model, or benign agents. It can only access publicly available registry information and register new agents through normal registration procedures. For its own registered agents, the attacker can freely design the corresponding skill descriptions, but cannot modify the descriptions of other agents.

\textbf{Attacker objective.}
The attacker aims to lure task routing requests originally intended for benign agents in a specific target domain, such as medicine, finance, law, or computer security, into their own hands. By crafting a deceptive skill description, the malicious agent seeks to obtain higher routing scores for target-domain queries and thus enter the selected agent set.

\textbf{Attacker constraints.}
The attacker does not directly manipulate the routing mechanism, compromise system components, or interfere with benign agents. Instead, it operates within normal IoA registration and execution rules, making the attack stealthy and difficult to distinguish from ordinary agent registration behavior.

\begin{figure}[t]
\centering{\includegraphics[width=0.45\textwidth]{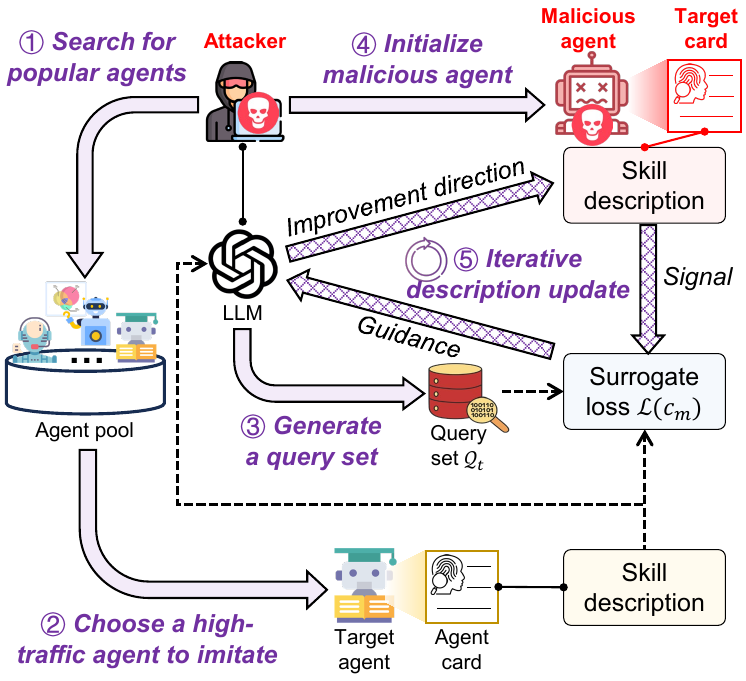}}
\caption{Workflow of SDD attack based on LLM techniques.}\label{fig2}
\end{figure}

\section{LLM-based Skill Describes Deceptive Attack}

\subsection{Deceptive Skill Description Generation}
\label{sec:sdd_attack}

As shown in Fig.~\ref{fig2}, we depict the workflow of an SDD attack in which a malicious agent $a_m$ aims to attract task routing requests originally intended for benign agents in a specific target domain. To this end, the malicious agent strategically crafts its skill description to appear highly relevant to the target domain. Different from conventional attacks that manipulate user prompts, model inputs, or communication messages, the proposed SDD attack exploits the semantic interface used for task routing in IoA systems.

\textbf{Step 1: Target Domain Identification and Skill Description Collection.}

In IoA systems, agents are typically registered with textual skill descriptions that summarize their capabilities, supported tasks, and service interfaces. Such descriptions are used to discover and select suitable agents for task routing, since forwarding all candidate agents' information to the LLM is impractical in large-scale IoA~\cite{lumer2025tool, fei2025mcp}. Therefore, an attacker can first identify a target domain $\mathcal{D}_t$, such as medicine, finance, law, or computer security, and collect the skill descriptions of benign agents associated with this domain.

Let $\mathcal{A}_t \subseteq \mathcal{A}$ denote the set of benign agents serving the target domain $\mathcal{D}_t$. Each benign agent $a_i \in \mathcal{A}_t$ is associated with a skill description $d_i$. The attacker obtains a set of target-domain skill descriptions as
\begin{equation}
\mathcal{D}_s^t = \{ d_i \mid a_i \in \mathcal{A}_t \},
\end{equation}
which serves as the semantic reference for constructing a deceptive malicious description. This step does not require access to the internal parameters of the router, since skill descriptions are naturally exposed or retrievable in IoA service discovery and agent routing workflows.

\textbf{Step 2: Query Set Construction via LLM Generation.}

To approximate the task distribution of the target domain, the attacker constructs a set of representative user queries
\begin{equation}
\mathcal{Q}_t = \{ q_1, q_2, \dots, q_K \}.
\end{equation}
Instead of manually designing these queries, the attacker leverages an LLM to generate semantically diverse and domain-consistent queries conditioned on the collected target-domain skill descriptions:
\begin{equation}
\mathcal{Q}_t \sim \mathcal{G}_{\mathrm{LLM}}(\mathcal{D}_s^t),
\end{equation}
where $\mathcal{G}_{\mathrm{LLM}}(\cdot)$ denotes the LLM-based query generation process. This enables the attacker to automatically approximate the types of user requests that are likely to be routed to benign agents in the target domain.

\textbf{Step 3: Iterative Deceptive Skill Description Optimization.}

Based on the generated query set $\mathcal{Q}_t$, the attacker optimizes the malicious skill description $d_m$ to increase the semantic relevance between the malicious agent and target-domain queries. The initial malicious description $d_m^{(0)}$ can be generated by summarizing the common capabilities of the target-domain agents:
\begin{equation}
d_m^{(0)} = \mathcal{S}_{\mathrm{LLM}}(\mathcal{D}_s^t),
\end{equation}
where $\mathcal{S}_{\mathrm{LLM}}(\cdot)$ denotes an LLM-based summarization or synthesis function.

At iteration $n$, the attack objective is to maximize the fraction of target-domain queries for which the malicious agent ranks higher than benign agents in the same domain:
\begin{equation}
\max_{d_m^{(n)}} \;
\frac{1}{|\mathcal{Q}_t|}
\sum_{q \in \mathcal{Q}_t}
\mathbb{I}
\left[
s(q, d_m^{(n)}) >
\max_{d_i \in \mathcal{D}_s^t} s(q, d_i)
\right],
\end{equation}
where $\mathbb{I}[\cdot]$ is the indicator function and $s(q,d)$ denotes the routing relevance score between query $q$ and skill description $d$.

Since the above objective is discrete and non-differentiable for black-box routing systems, we use a surrogate optimization objective:
\begin{equation}
\mathcal{L}(d_m) =
\sum_{q \in \mathcal{Q}_t}
\ell
\left(
\max_{d_i \in \mathcal{D}_s^t} s(q,d_i)
-
s(q,d_m)
\right),
\end{equation}
where $\ell(\cdot)$ can be a margin-based or logistic loss function. The malicious skill description is then updated through LLM-guided semantic rewriting:
\begin{equation}
d_m^{(n+1)}
=
\mathcal{U}_{\mathrm{LLM}}
\left(
d_m^{(n)}, \mathcal{Q}_t, \mathcal{L}(d_m^{(n)})
\right),
\end{equation}
where $\mathcal{U}_{\mathrm{LLM}}(\cdot)$ represents an LLM-driven rewriting operator. The rewriting process aims to preserve the apparent legitimacy of the malicious agent while improving its semantic alignment with target-domain queries.

Through iterative optimization, the malicious skill description gradually shifts toward a semantic representation that is more competitive than benign domain agents under the routing mechanism. As a result, the malicious agent becomes more likely to be selected during task routing without modifying the router, compromising existing agents, or interfering with the underlying communication infrastructure.

\subsection{Semantic Routing Manipulation}
\label{sec:routing_manipulation}

The proposed SDD attack manipulates routing outcomes by changing only the malicious agent's skill description. Given a target-domain query $q \in \mathcal{Q}_t$, the router ranks candidate agents according to the semantic relevance between $q$ and their skill descriptions. The malicious agent succeeds when its optimized description obtains a higher relevance score than benign descriptions in the target domain:
\begin{equation}
s(q,d_m) >
\max_{d_i \in \mathcal{D}_s^t} s(q,d_i).
\end{equation}

We define the routing dominance ratio of the malicious agent as
\begin{equation}
\rho_m =
\frac{1}{|\mathcal{Q}_t|}
\sum_{q \in \mathcal{Q}_t}
\mathbb{I}
\left[
a^*(q)=a_m
\right],
\end{equation}
where $a^*(q)$ denotes the agent selected by the router for query $q$. This metric quantifies the proportion of target-domain queries hijacked by the malicious agent.

For top-$K$ routing, the attack is considered successful if the malicious agent appears in the selected candidate set $\mathcal{A}_q^K$:
\begin{equation}
\rho_m^{K} =
\frac{1}{|\mathcal{Q}_t|}
\sum_{q \in \mathcal{Q}_t}
\mathbb{I}
\left[
a_m \in \mathcal{A}_q^K
\right].
\end{equation}
This metric reflects the practical risk that malicious agents may enter the execution chain even when the router delegates a task to multiple candidate agents.

\section{Experiment Results}
\begin{figure*}[t]
    \centering

    \begin{subfigure}[b]{0.32\textwidth}
        \centering
        \includegraphics[width=\textwidth]{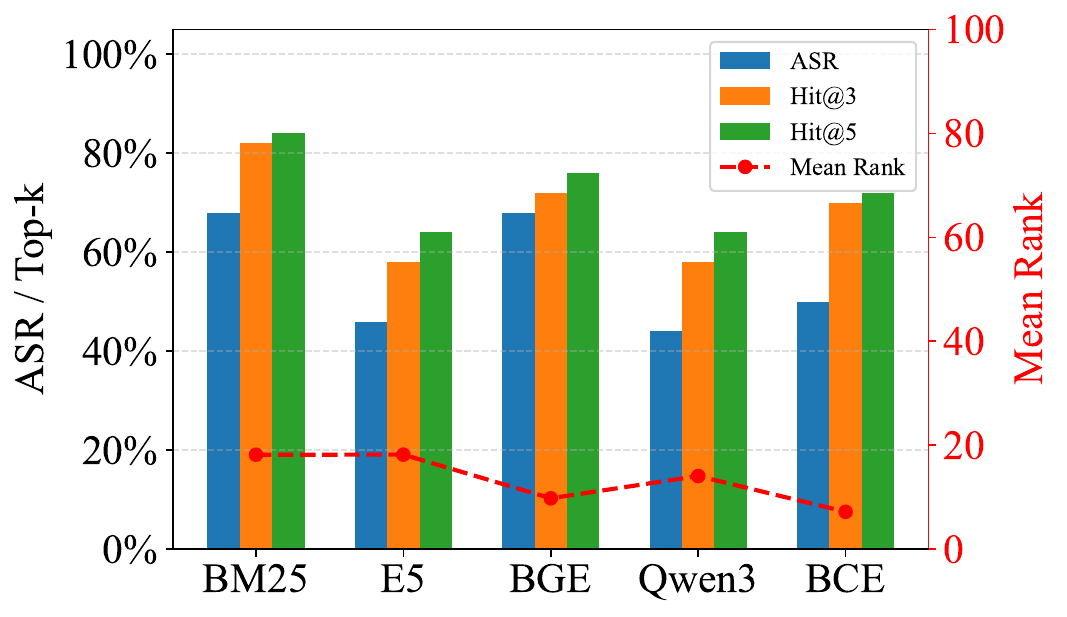}
        \caption{Math}
        \label{fig:math}
    \end{subfigure}
    \hfill
    \begin{subfigure}[b]{0.32\textwidth}
        \centering
        \includegraphics[width=\textwidth]{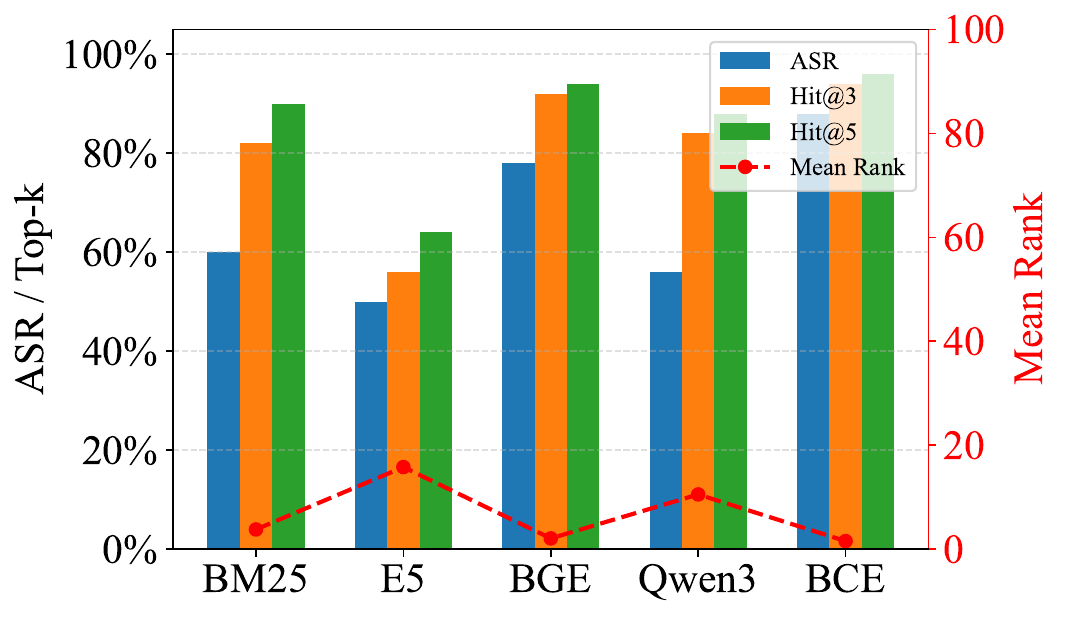}
        \caption{Physics}
        \label{fig:physics}
    \end{subfigure}
    \hfill
    \begin{subfigure}[b]{0.32\textwidth}
        \centering
        \includegraphics[width=\textwidth]{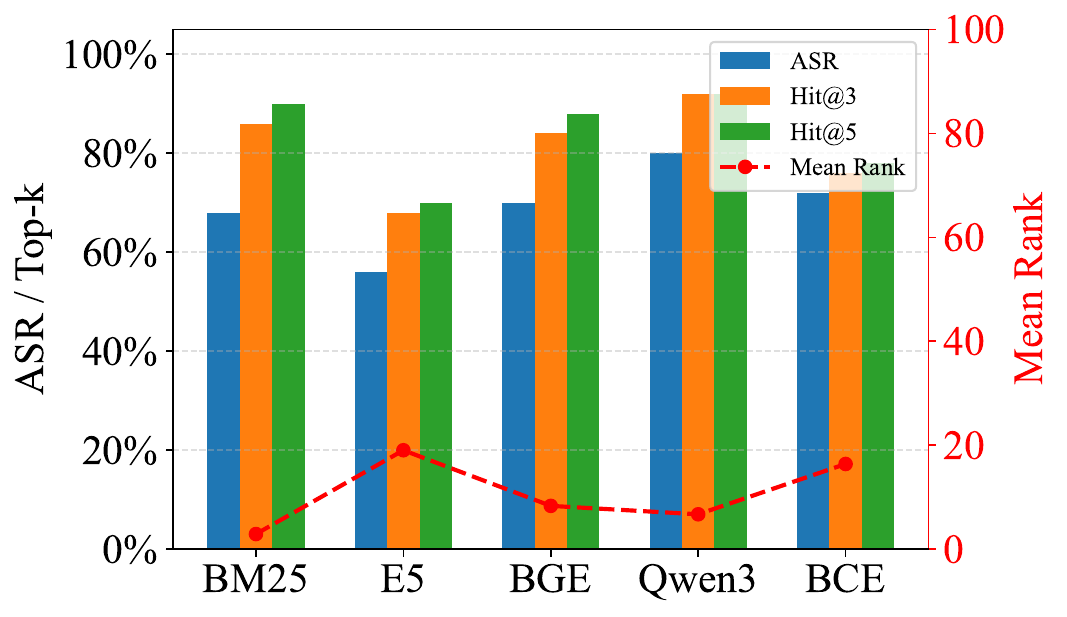}
        \caption{Finance}
        \label{fig:finance}
    \end{subfigure}

    \vspace{0.6em}

    \begin{subfigure}[b]{0.32\textwidth}
        \centering
        \includegraphics[width=\textwidth]{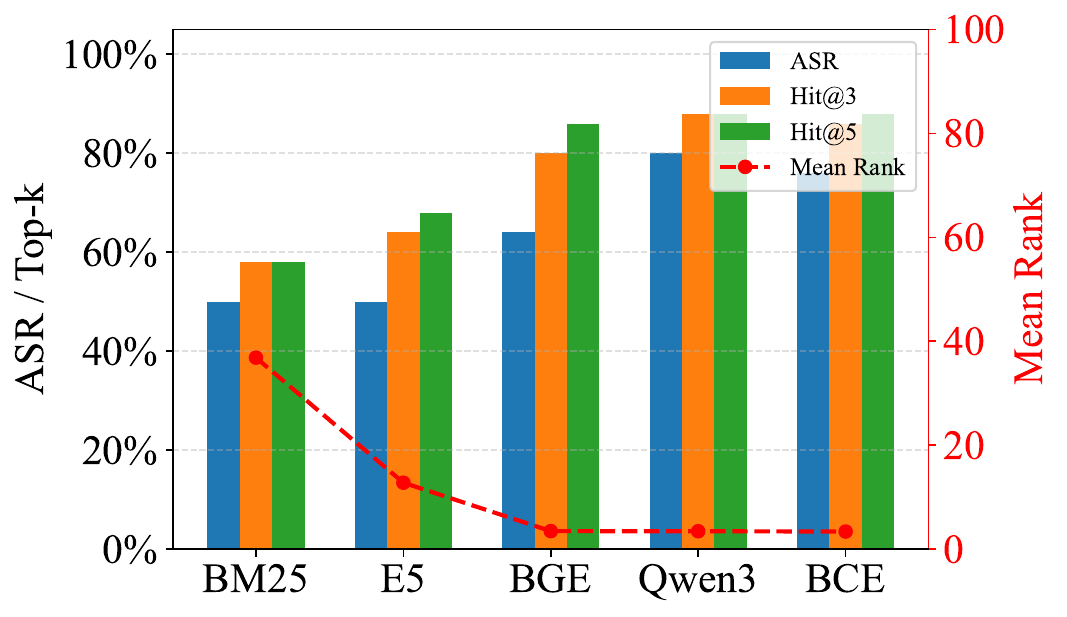}
        \caption{Computer Science}
        \label{fig:cs}
    \end{subfigure}
    \hfill
    \begin{subfigure}[b]{0.32\textwidth}
        \centering
        \includegraphics[width=\textwidth]{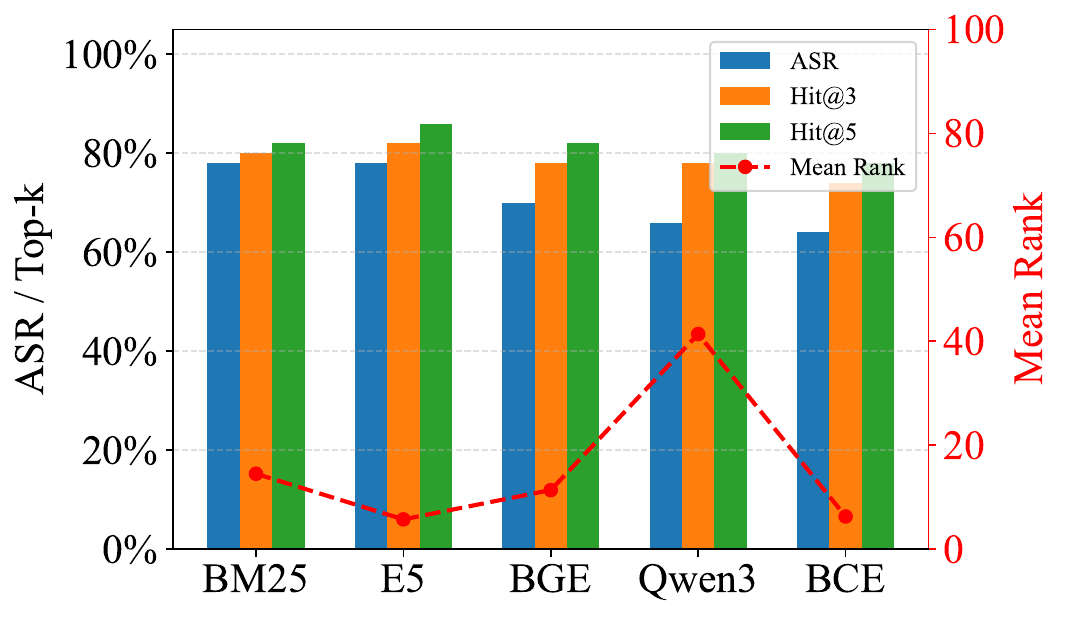}
        \caption{Law}
        \label{fig:law}
    \end{subfigure}
    \hfill
    \begin{subfigure}[b]{0.32\textwidth}
        \centering
        \includegraphics[width=\textwidth]{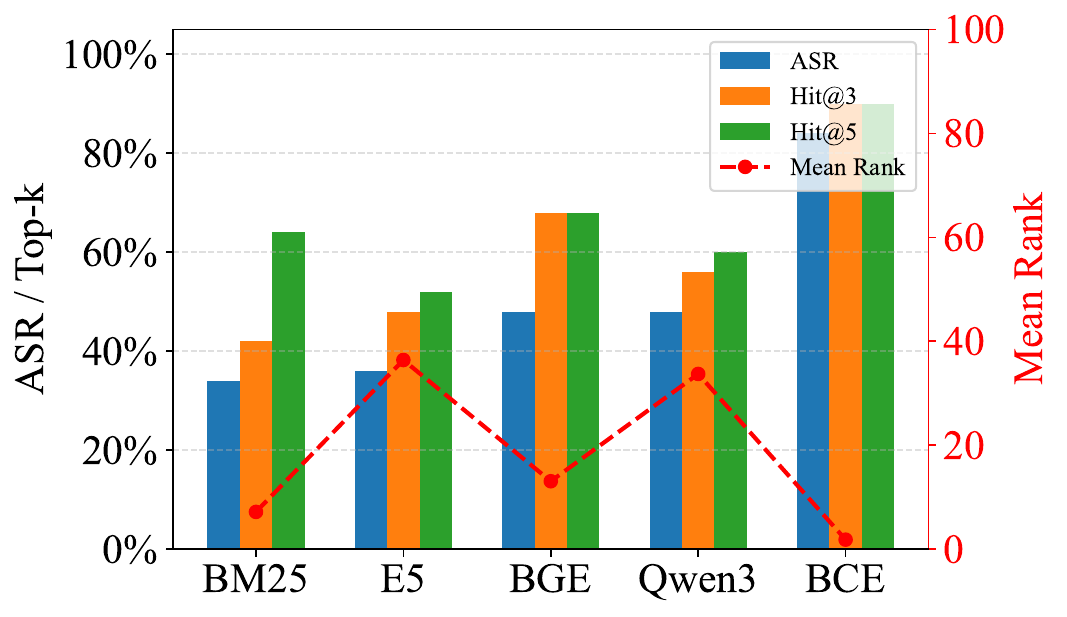}
        \caption{Medicine}
        \label{fig:medicine}
    \end{subfigure}

    \vspace{0.6em}

    \begin{subfigure}[b]{0.32\textwidth}
        \centering
        \includegraphics[width=\textwidth]{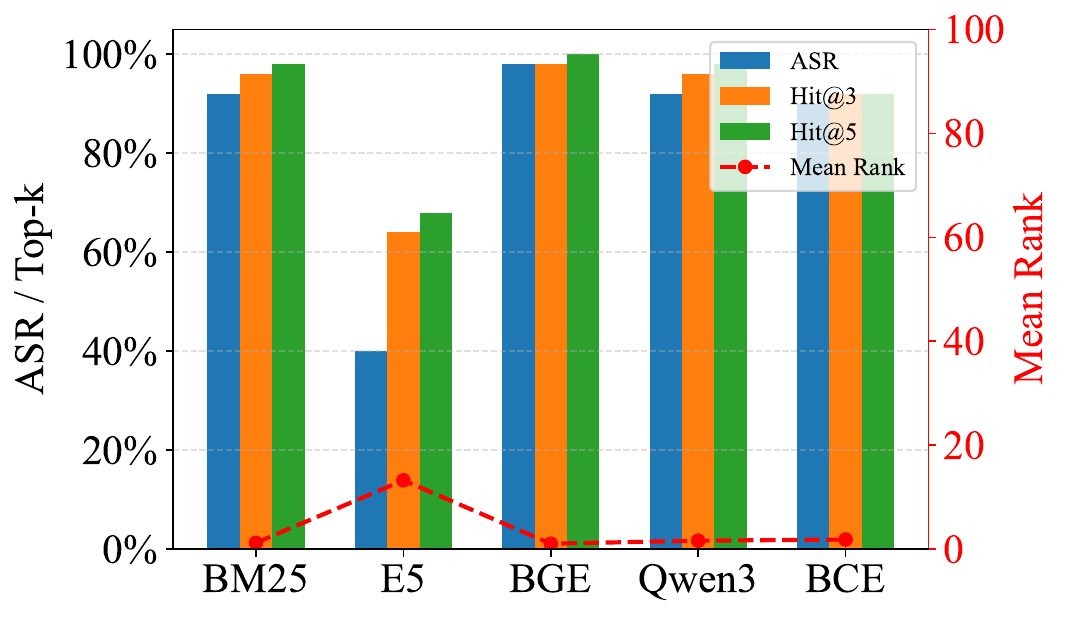}
        \caption{Algebra}
        \label{fig:algebra}
    \end{subfigure}
    \hfill
    \begin{subfigure}[b]{0.32\textwidth}
        \centering
        \includegraphics[width=\textwidth]{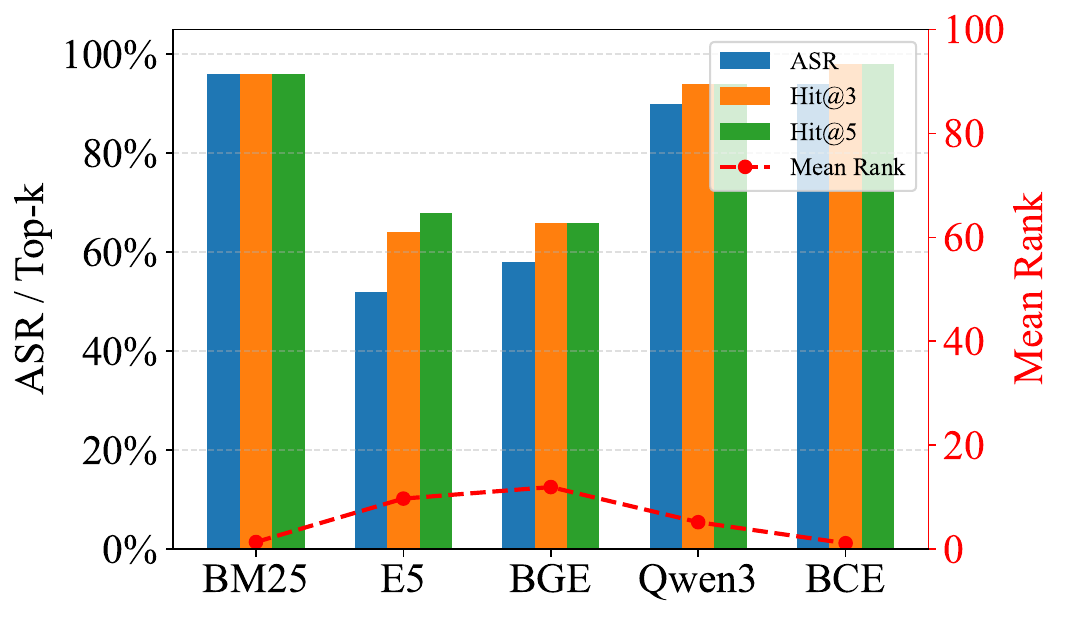}
        \caption{History}
        \label{fig:history}
    \end{subfigure}
    \hfill
    \begin{subfigure}[b]{0.32\textwidth}
        \centering
        \includegraphics[width=\textwidth]{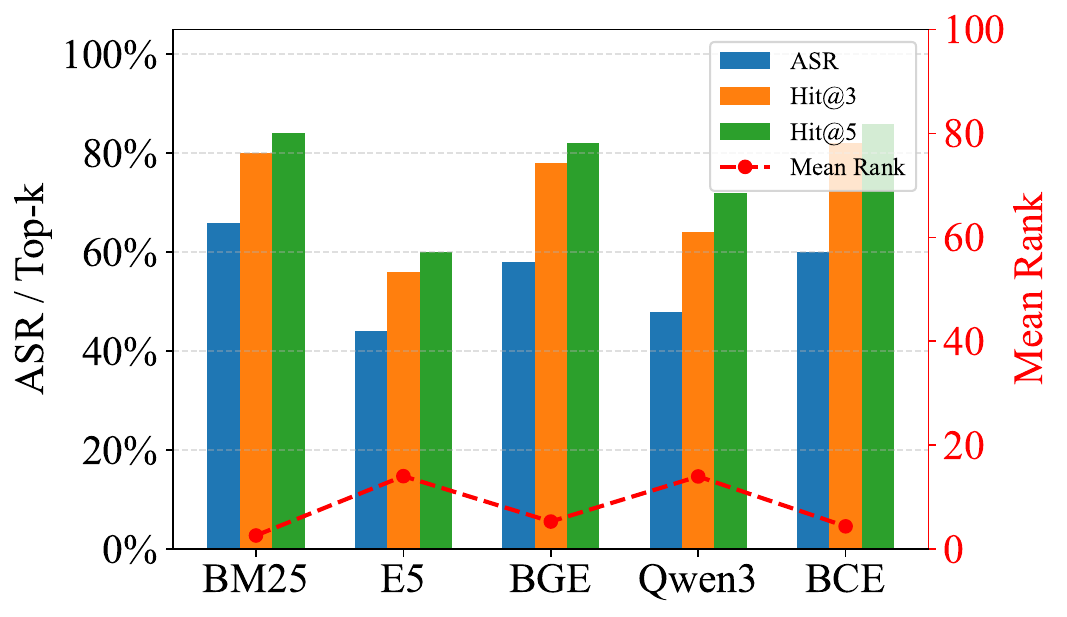}
        \caption{Computer Science}
        \label{fig:computer_science}
    \end{subfigure}

    \caption{Comparison of attack performance across nine domains, including ASR, Hit@3, Hit@5, and mean rank under different retrieval methods.}
    \label{fig:nine_domains}
\end{figure*}

\subsection{Experimental Setup}

\textbf{IoA Environment.}
 We construct the agent pool based on the LiveMCPBench dataset~\cite{mo2025livemcpbench}, which contains 527 real-world tools/agents collected from MCP servers. Following the A2A protocol~\cite{A2A}, each agent/tool is associated with a natural language skill description that specifies its functionality and supported tasks. These agents constitute the benign agent pool in our simulated IoA environment. To evaluate the proposed attack, we additionally register one malicious agent into the agent pool, whose skill description is generated or manipulated according to different attack strategies. Therefore, each routing experiment involves 528 candidate agents in total, including 527 benign agents/tools and one malicious agent.

\textbf{Task Set.}
We use the Massive Multitask Language Understanding (MMLU) benchmark~\cite{mmlu} to construct the task set. Specifically, we select nine representative domains, including math, physics, finance, computer security, law, medicine, algebra, history, and computer science. For each domain, we randomly sample 100 tasks, resulting in 900 user queries in total. Each domain is treated as a target routing scenario, where the malicious agent attempts to attract queries that should be routed to benign domain-relevant agents.

\textbf{Routing Mechanism.}
We evaluate the attack under five representative retrieval and embedding-based routing mechanisms: BM25~\cite{bm25}, E5~\cite{e5}, BGE~\cite{bge}, Qwen Embedding 8B~\cite{zhang2025qwen3embedding}, and BCE~\cite{youdao_bcembedding_2023}. BM25 is a sparse lexical retrieval method based on term matching, while E5, BGE, Qwen Embedding 8B, and BCE are dense embedding models that encode both user queries and agent skill descriptions into semantic vector representations. For each query $q$ and agent skill description $d_i$, we compute their relevance score after retrieval or embedding. For dense embedding methods, the routing score is calculated by cosine similarity:
\begin{equation}
s(q,d_i)=
\frac{\mathbf{e}_q^\top \mathbf{e}_i}
{\|\mathbf{e}_q\|\|\mathbf{e}_i\|},
\end{equation}
where $\mathbf{e}_q$ and $\mathbf{e}_i$ denote the embeddings of the user query and the skill description, respectively. For BM25, the BM25 relevance score is directly used for ranking. All candidate agents are ranked according to their relevance scores, and the ranking position of the malicious agent is recorded for evaluation.

\textbf{Metrics.}
We adopt three metrics to evaluate the effectiveness of the proposed attack, enumerated as follows.

1) \emph{Attack Success Rate (ASR)} measures the proportion of queries for which the malicious agent is ranked first by the routing mechanism:
\begin{equation}
\mathrm{ASR} =
\frac{1}{|\mathcal{Q}|}
\sum_{q \in \mathcal{Q}}
\mathbb{I}
\left[
r_m(q)=1
\right],
\end{equation}
where $\mathcal{Q}$ is the evaluation query set, while $r_m(q)$ denotes the ranking position of the malicious agent for query $q$.

2) \emph{Top-$K$ Hit Rate} measures the proportion of queries for which the malicious agent appears within the top-$K$ ranked candidates:
\begin{equation}
\mathrm{Hit@}K =
\frac{1}{|\mathcal{Q}|}
\sum_{q \in \mathcal{Q}}
\mathbb{I}
\left[
r_m(q) \leq K
\right].
\end{equation}
In our experiments, we report Hit@3 and Hit@5 to evaluate the risk that the malicious agent being included in the candidate task execution set.

3) \emph{Mean Rank (MR)} measures the average ranking position of the malicious agent across all evaluation queries:
\begin{equation}
\mathrm{MR} =
\frac{1}{|\mathcal{Q}|}
\sum_{q \in \mathcal{Q}}
r_m(q).
\end{equation}
A lower MR indicates that the malicious agent is ranked closer to the top and therefore has a higher chance of being selected during task routing.

\subsection{Numerical Results and Analysis}
\begin{table}[t]
\centering
\caption{Comparison of different attacks in terms of ASR.}
\label{tab:attack_comparison}
\small
\setlength{\tabcolsep}{3.5pt}
\begin{tabular}{lccccc}
\toprule
\textbf{Router} & \textbf{Exag.} & \textbf{Keyword} & \textbf{Gen.} & \textbf{Imperson.} & \textbf{SDD} \\
\midrule
BM25  & 6.67\%  & 10.00\% & 0.00\% & 4.22\%  & \textbf{68.00\%} \\
E5    & 11.11\% & 26.22\% & 1.11\% & 10.22\% & \textbf{50.22\%} \\
BGE   & 44.44\% & 58.22\% & 3.33\% & 40.00\% & \textbf{68.00\%} \\
Qwen3 & 31.11\% & 30.67\% & 0.00\% & 24.44\% & \textbf{67.11\%} \\
BCE   & 38.67\% & 38.89\% & 1.11\% & 23.78\% & \textbf{75.33\%} \\
\bottomrule
\end{tabular}
\end{table}

Fig.~\ref{fig:nine_domains} shows the routing results across nine task domains and five routing mechanisms. Overall, the results demonstrate that IoA systems relying on semantic task routing are highly vulnerable to skill description manipulation. Although the agent pool contains 527 benign agents/tools and only one malicious agent, the malicious agent can still be frequently ranked at the top by manipulating its skill description. In particular, the SDD attack achieves up to 98\% ASR in the algebra domain under BGE, indicating that a single malicious agent can almost always dominate benign candidates in certain domain-specific routing scenarios. 

The vulnerability is consistently observed across all evaluated routing mechanisms, including BM25, E5, BGE, Qwen3, and BCE. This suggests that the vulnerability does not originate from a particular retrieval or embedding model, but from the general dependence of task routing on semantic relevance between user queries and self-declared skill descriptions. Moreover, Hit@3 and Hit@5 are generally higher than ASR, meaning that even when the malicious agent is not ranked first, it can still frequently appear among the top candidate agents and enter the service chain. The low mean-rank values in several settings further indicate that the malicious agent is often placed very close to the top of the routing list. These results reveal that semantic relevance alone is insufficient for trustworthy task routing, and that domain-specific deceptive descriptions can significantly bias routing decisions in open IoA systems.

Table~\ref{tab:attack_comparison} compares SDD with four heuristic skill description manipulation strategies: capability exaggeration (Exag.), keyword stuffing (Keyword), generic domain description (Gen.), and agent impersonation (Imperson.). The results show that heuristic manipulation strategies only expose the routing vulnerability in an unstable manner. Gen. is almost ineffective across all routers, indicating that overly broad descriptions cannot reliably attract domain-specific routing requests. Keyword generally performs better than Exag. and Imperson. under several routers, such as E5 and BGE, suggesting that keyword-level relevance can influence semantic matching. However, its effectiveness varies considerably across routing mechanisms and remains much lower than SDD in most cases. Imperson. also achieves moderate ASR under BGE and Qwen3, but fails to provide consistent routing bias across all routers.

In contrast, SDD consistently achieves the highest ASR across all evaluated routing mechanisms, including BM25, E5, BGE, Qwen3, and BCE. This indicates that the major security risk does not simply arise from adding domain keywords, exaggerating capabilities, or imitating benign agents. Instead, SDD attack can better align with the semantic routing space in an adaptive and domain-specific manner. These results suggest that IoA systems should not rely solely on textual relevance for task routing, since well-crafted skill descriptions may appear semantically legitimate while still causing unsafe delegation decisions.

The experimental results reveal a fundamental weakness in current IoA routing designs: semantic relevance alone is insufficient to ensure trustworthy task delegation. A malicious agent can exploit the gap between declared skills and actual capabilities to obtain favorable routing positions, thereby increasing the risk of low-quality, misleading, or harmful service execution. This highlights the need for trust-aware routing, capability verification, and anomaly detection mechanisms in future IoA systems.

\section{Conclusion}
This paper reveals a new attack surface in IoA systems, where malicious agents can manipulate self-declared skill descriptions to bias semantic matching-based task routing decisions. We formalize this kind of attack as the Skill Description Deception (SDD) attack and show that it can achieve an attack success rate of up to 98\%, demonstrating the vulnerability of semantic routing mechanisms in open agentic network environments. 
These findings highlight the need for trustworthy routing mechanisms in future IoA systems. Promising directions include deceptive-description detection, abnormal routing behavior analysis, capability verification, trust-aware routing, and adaptive defense mechanisms against malicious agents.

\bibliographystyle{IEEEtran}
\bibliography{ref}

\end{document}